\documentclass[10pt,conference]{IEEEtran}
\IEEEoverridecommandlockouts
\usepackage{cite}
\usepackage{amssymb}
\usepackage{amsfonts}
\usepackage{graphicx}
\usepackage{amsmath}
\usepackage{amsthm}
\usepackage{dsfont}
\usepackage{url}
\usepackage{textcomp}
\usepackage{xcolor}
\usepackage{algorithm}
\usepackage{algpseudocode}
\usepackage{tcolorbox}
\usepackage{pgfplots}
\usepackage{listings}
\usepackage{svg}
\pgfplotsset{compat=1.17} 
    
\begin{document}

\title{Tensor Decompositions and Adiabatic Quantum Computing for Discovering Practical Matrix Multiplication Algorithms}

\author{
\IEEEauthorblockN{Valter Uotila}
\IEEEauthorblockA{
University of Helsinki\\
valter.uotila@helsinki.fi}
}

\maketitle

\begin{abstract}
Quantum computing and modern tensor-based computing have a strong connection, which is especially demonstrated by simulating quantum computations with tensor networks. The other direction is less studied: quantum computing is not often applied to tensor-based problems. Considering tensor decompositions, we focus on discovering practical matrix multiplication algorithms and develop two algorithms to compute decompositions on quantum computers. The algorithms are expressed as higher-order unconstrained binary optimization (HUBO) problems, which are translated into quadratic unconstrained binary optimization (QUBO) problems. Our first algorithm is decompositional to keep the optimization problem feasible for the current quantum devices. Starting from a suitable initial point, the algorithm discovers tensor decomposition corresponding to the famous Strassen matrix multiplication algorithm, utilizing the current quantum annealers. Since the decompositional algorithm does not guarantee minimal length for found tensor decompositions, we develop a holistic algorithm that can find fixed-length decompositions. Theoretically, by fixing a shorter length than the length for the best-known decomposition, we can ensure that the solution to the holistic optimization problem would yield faster matrix multiplication algorithms.
\end{abstract}


\section{Introduction}

Quantum computing is rapidly evolving and showing promising advances in hardware development. Based on the quantum computing vendors' roadmaps, we will have access to more scalable quantum computing systems in the near future \cite{Hellstrom_20220908, GoogleQuantumApproach, IBMQuantumRoadmap}. Concurrently, there is a growing need for innovative quantum algorithms and finding suitable real-world problems that can be solved on the current and near-term quantum computing devices. Moreover, quantum computers and algorithms should show certain advantages and utilities over the best-known classical solutions. Demonstrating the usefulness and superiority of quantum computing compared to the best-known classical algorithms is notoriously challenging \cite{kim_2023}.

In this work, we focus on developing two algorithms for computing tensor decompositions from a quantum computational perspective. Although tensors are at the heart of quantum computing and various tensor-based computational methods have been proposed to model and speed up quantum simulations \cite{Orus2018Tensor, Huggins2018Towards, Gray2020Hyper-optimized, kim_2023, Tindall_Fishman_Stoudenmire_Sels_2024}, the other direction has been studied less. Quantum computing has not been widely applied to tensor-based problems or calculating tensor decompositions. We especially focus on discovering matrix multiplication algorithms since they can be expressed as tensor decompositions. 

Our work is inspired and motivated by the recent discoveries made in \cite{Fawzi_Balog_Huang_Hubert_Romera_Paredes_Barekatain_Novikov_RRuiz_Schrittwieser_Swirszcz_et_2022}. They have developed a deep reinforcement-based machine learning approach, AlphaTensor, for automatic discovery of matrix multiplication algorithms. AlphaTensor was able to find matrix multiplication algorithms, i.e., tensor decompositions, that outperform the state-of-the-art complexity of many currently best algorithms. A particular highlight of this work was the case of multiplying $4 \times 4$ matrices in a finite field $\mathds{F}_2$, where AlphaTensor's algorithm improves on Strassen's two-level algorithm.

Discovering faster matrix multiplication algorithms is a well-motivated but extremely hard real-life problem. The matrix multiplications are a fundamental operation for countless classical algorithms, especially in machine learning. Even small improvements in these algorithms would lead to significant practical savings in computational resource utilization. On the other hand, the problem's hardness is demonstrated by the fact that it is not even known how to multiply $3 \times 3$ matrices in the most efficient way \cite{Fawzi_Balog_Huang_Hubert_Romera_Paredes_Barekatain_Novikov_RRuiz_Schrittwieser_Swirszcz_et_2022}. Our work aims to find practical matrix multiplication algorithms instead of focusing on theoretical bounds for the exponent of the optimal time complexity of matrix multiplication which has also advanced remarkably recently \cite{Strassen_1986,Duan_Wu_Zhou_2023,Williams_Xu_Xu_Zhou_2024}. 

In this work, we develop two different quantum computing algorithms for computing tensor decompositions. We call the first algorithm decompositional because it divides the hard problem into steps that are solved sequentially. Since the decompositional algorithm's objective is not to minimize the total length of the decomposition, which is the exact measure of the performance of a matrix multiplication algorithm, we propose a second algorithm called a holistic algorithm. The holistic algorithm aims to find fixed-length tensor decompositions. Both of the algorithms utilize a higher-order unconstrained binary optimization (HUBO) model, which is translated into a quadratic unconstrained binary optimization (QUBO) model. From the algorithm design perspective, our approach aligns with \cite{Gaitan}, where the authors developed an adiabatic algorithm for the graph isomorphism problem.

We implement both algorithms utilizing D-wave's Ocean software \cite{ocean}. We encode the problems utilizing binary quadratic and binary polynomial instances. Furthermore, we have integrated classical CPLEX and Gurobi solvers as part of the framework. While we aim to be mathematically precise, we present the concrete algorithms and approach the topic with concrete examples. The implementation is available on GitHub \cite{uotila_2023}.

The key contributions of this paper are outlined as follows:
\begin{itemize}
    \item We develop decompositional and holistic algorithms to compute tensor decompositions, especially for discovering matrix multiplication algorithms. 
    \item Given a good initial point for the decompositional algorithm, we are able to discover the famous, provably optimal Strassen matrix multiplication algorithm using quantum annealers or classical simulators and solvers.
    \item We show how the holistic algorithm can be technically solved using quantum annealers and classical solvers, and we study its characteristics around the known optimal points, which correspond to the best-known matrix multiplication algorithms.
\end{itemize}

The paper is organized so that we first briefly introduce the basics of adiabatic quantum computing and quantum annealing with a focus on rewriting higher-degree unconstrained binary optimization (HUBO) problems as quadratic unconstrained binary optimization (QUBO) problems. Then, we review the theory of computing tensor decompositions with the example of multiplying $2 \times 2$ matrices. After that, we represent the main algorithms. We describe their implementation and present the results, e.g., how we are able to discover Strassen's matrix multiplication algorithm. Finally, we conclude the paper with a discussion and propose future work.
\section{Preliminaries}

\subsection{Adiabatic quantum computing and quantum annealing}

Adiabatic quantum computing \cite{Farhi_Goldstone_Gutmann_Sipser_2000, adiabatic} is a quantum computing paradigm based on the adiabatic theorem. Quantum annealing is a special branch of adiabatic quantum computing utilizing quantum annealers. While adiabatic quantum computing has shown to be polynomially equivalent to the quantum circuit model \cite{Aharonov_2007}, the computational power of quantum annealing is not precisely understood \cite{Mukherjee_Chakrabarti_2015}. In practice, these devices show promising results in quantum simulation \cite{dwave_supremacy}.

Both adiabatic quantum computing and quantum annealing are based on the adiabatic theorem. Initially, a complex Hamiltonian is constructed so that its unknown ground state represents a solution to the given problem. Subsequently, a system is set up with a simple Hamiltonian and initialized to its ground state that is easy to identify. The simple Hamiltonian is gradually involved in the more complex, pre-identified Hamiltonian. Following the principles of the adiabatic theorem, the system consistently stays in the ground state throughout this evolution. At the end of this process, the system's ground state provides the solution to the given problem.

\subsection{Higher-order and quadratic unconstrained binary optimization problems}\label{subsection:hubo}

Quadratic Unconstrained Binary Optimization (QUBO) problems \cite{lucas_2014} are a common formalism to express problems for quantum annealers. Let $\mathds{B} = \left\{0,1\right\}$ be the set of binary values and $n > 0$. The set $\mathds{B}^n$ denotes the set of all binary bit strings of length $n$. The QUBO problem
comprises minimizing the objective function $f \colon \mathds{B}^n \to \mathds{R}$ defined by
\begin{equation}\label{eq:qubo}
    f(x) = x^{\top}Qx = \sum_{i = 1}^{n}\sum_{j = i}^{n} q_{i,j}x_{i}x_{j},
\end{equation}
where $Q$ is a real-valued symmetric $n \times n$ matrix with elements $q_{i,j}$ for $i,j = 1, \ldots, n$. The variable $x$ is a binary-valued column vector of length $n$, and the variable $x^{\top}$ is the corresponding binary-valued row vector.

In this work, we also apply a high-order (high-degree) unconstrained binary optimization (HUBO) model \cite{boros_2002}. The HUBO problems are defined similarly to QUBO problems except that we allow higher degree interactions between the binary variables. Formally this means that for a binary variable vector $x \in \mathds{B}^n$ of length $n$ and for the index set $V = \left\{1, \ldots, n \right\}$ the objective function is
\begin{equation}\label{eq:hubo_formal}
    f(x) = \sum_{S \subset V}c_{S}\prod_{i \in S}x_{i},
\end{equation}
where $c_{S} \in \mathds{R}$ and $S$ runs over all the subsets of $V$. If the variable corresponding to some subset does not appear in the objective function, its coefficient is $0$. The HUBO model subsumes the QUBO model when we restrict the size of the set $S$ to be either 1 (linear terms) or 2 (quadratic terms).

Besides quantum annealers, the QUBO type of problems can be solved in various ways \cite{hermanni}. On the circuit-based universal quantum computers, we can apply the Quantum Approximate Optimization Algorithm (QAOA) \cite{farhi2014quantum} or Variational Quantum Eigensolver (VQE) \cite{Peruzzo_2014} to find the ground state of a Hamiltonian which is created for the optimization problem. Additionally, the neutral-atom quantum computing architectures \cite{Brennen1998QUANTUM} implement methods to map QUBOs to problems that are natively supported on these devices \cite{unitdiskmapping}. Moreover, we can solve QUBOs using simulated annealing and classical solvers such as Gurobi and CPLEX.

Both QUBO \eqref{eq:qubo} and HUBO \eqref{eq:hubo_formal} are NP-hard problems \cite{lucas_2014, boros_2002}. HUBO is also hard in practice and could be approximately solved with non-convex optimizers. Next we describe how to rewrite HUBO problems as QUBO problems using auxiliary variables. In order to utilize the current quantum annealers, we apply the Ocean framework's utility of translating HUBO problems into QUBO problems. The translation is based on a polynomial reduction by minimum selection or by substitution \cite{polynomial_reductions}. We rely on the automatic functionality in the Ocean framework named \texttt{make\_quadratic}.

The HUBO to QUBO reduction based on minimum selection \cite{polynomial_reductions} is based on the following identity
\begin{equation*}
    xyz = \max_{w}w(x+y+z-2),
\end{equation*}
which iteratively replaces the higher order terms $xyz$ with lower order terms by introducing additional variables $w$. Depending on the order of this replacement process, the QUBO formula can have different formats and different numbers of binary variables.

Rewriting by substitution utilizes the following formula
\begin{equation*}
    xyz = \min_{w}\left\{wz + \mathrm{MP}(x, y ; w)\right\},
\end{equation*}
where $M > 1$  is a penalty weight and $P$ is a penalty function that has the lowest value when $w = xy$. The details of why these rewriting methods reach the same minimum are explained in \cite{polynomial_reductions}.

Since optimizing HUBO problems is often not supported by classical or quantum solvers, we translate the problem into QUBO using the previously described scheme. On the other hand, circuit-model-based systems and universal quantum computers are able to encode HUBO problems without the transformation into QUBO~\cite{10313783, 10.1145/3478519}. This is a crucial practical difference between universal quantum computing and quantum annealing, which supports only quadratic interactions. For example, in Fig. (\ref{fig:Hamiltonian}), we represent the Pennylane \cite{bergholm2022pennylane} implementation of mapping HUBO into a cost Hamiltonian, which can be used in QAOA. Unfortunately, the size of the problems in our work is too large for methods in universal quantum computing.

\begin{figure}[tbp]
    \centering
    \begin{lstlisting}[language=Python]
    vars = HUBO.get_variables()
    vars_to_qubits = dict(zip(vars, range(n_qubits)))
    coeffs, obs = [], []
    
    # HUBO.get_terms() is a dictionary storing
    # { (var1, var2, ..., varn): coeff }
    for variable_tuple, coeff in HUBO.get_terms():
        coeffs.append(coeff)
        pauli_list = []
        for var in variable_tuple:
            pauliZ = qml.PauliZ(vars_to_qubits[var])
            pauli_list.append(pauliZ)
        obs.append(qml.operation.Tensor(*pauli_list))
        
    cost_Hamiltonian = qml.Hamiltonian(coeffs, obs)
    \end{lstlisting}
    \caption{Encoding HUBO into Hamiltonian with Pennylane}
    \label{fig:Hamiltonian}
\end{figure}

\subsection{Strassen algorithm to multiply $2 \times 2$ matrices}\label{subsection:strasse}

In this part, we demonstrate how $2 \times 2$ matrix multiplication algorithms can be calculated faster with the Strassen algorithm \cite{Strassen_1969}. Following matrix multiplication is the standard algorithm for multiplying $2 \times 2$ matrices:
\begin{align}\label{eq:2x2}
\begin{bmatrix}
a & b \\
c & d
\end{bmatrix}\cdot\begin{bmatrix}
e & f \\
g & h
\end{bmatrix} &= \begin{bmatrix}
a\cdot e + b \cdot g & a \cdot f + b \cdot h \\
c \cdot e + d\cdot g & c\cdot f + d\cdot h
\end{bmatrix}.
\end{align}
The number of scalar multiplications ($\cdot$) is a concrete measure to determine when one matrix multiplication algorithm is better than the other. In the standard algorithm for multiplying $2 \times 2$ matrices, we perform eight scalar multiplications. The well-known fact is that the provably optimal number of multiplications is seven, given by the Strassen algorithm \cite{Strassen_1969}. Let the two matrices be the same as in standard matrix multiplication (\ref{eq:2x2}). Following the Strassen algorithm, we calculate the following intermediate values
\begin{align*}
    m_1 &= (a + d) \cdot (e + h) \quad  &m_4 = d \cdot (g - e) \\
    m_2 &= (c + d) \cdot e \quad  &m_5 = (a + b) \cdot h \\
    m_3 &= a \cdot (f - h) \quad &m_6 = (c - a) \cdot (e + f) \\
    m_7 &= (b - d) \cdot (g + h).
\end{align*}
Now, we have used only seven scalar multiplications to calculate values $m_i$ for $i = 1, \ldots, 7$. The Strassen algorithm shows that the final result of the matrix multiplication is
\begin{align*}
\begin{bmatrix}
a & b \\
c & d
\end{bmatrix}&\begin{bmatrix}
e & f \\
g & h
\end{bmatrix} \\ &= 
\begin{bmatrix}
    m_1 + m_4 - m_5 + m_7 & m_3 + m_5 \\
    m_2 + m_4 & m_1 - m_2 + m_3 + m_6
\end{bmatrix}.
\end{align*}

\subsection{Tensor decompositions and matrix multiplication algorithms}

This part follows the problem formulation presented in \cite{Fawzi_Balog_Huang_Hubert_Romera_Paredes_Barekatain_Novikov_RRuiz_Schrittwieser_Swirszcz_et_2022}. More detailed descriptions can be found in \cite{Burgisser_Clausen_Shokrollahi_1997,Landsberg_2017,Lim_2021}. We focus on tensor decomposition to find practical matrix multiplication algorithms. We consider multiplying a general $n \times m$ matrix with $m \times p$ matrix so that their terms are either in field $\mathds{R}$ or $\mathds{F}_2 = \left\{0, 1 \right\}$. By definition, matrix multiplication is a bilinear function
\begin{displaymath}
M_{n,m,p} \colon \mathds{R}^{n \times m} \times \mathds{R}^{m\times p} \to \mathds{R}^{n \times p}, \quad M_{n,m,p}(A, B) = AB.
\end{displaymath}
This bilinear operator is defined by a tensor (cf. linear maps are defined by matrices). Now, the core question is to find a tensor decomposition which represents this bilinear operator
\begin{equation}\label{eq:decomposition}
    M_{n,m,p}(A, B) = \sum_{i = 1}^{R} f_i(A)g_i(B)W_i,
\end{equation}
with linear mappings $f_i \colon \mathds{R}^{n \times m} \to \mathds{R}$ and $g_i \colon \mathds{R}^{m \times p} \to \mathds{R}$ and $W_i \in \mathds{R}^{n \times p}$. The value $R$ in the sum gives the number of required scalar multiplications (for example, seven in the case of the Strassen algorithm) to perform the matrix multiplication. The value $R$ is also called a rank of the decomposition. Any valid decomposition (\ref{eq:decomposition}) provides a method to multiply $n \times m$ matrix with $m \times p$ matrix. The difficulty of the problem is to find a decomposition whose rank is minimal. This problem, along with many other tensor decomposition problems, is proved to be NP-hard \cite{hillar_2013}. The standard matrix multiplication algorithm, such as \eqref{eq:2x2}, has a tensor representation, which can be calculated with Alg.~ \ref{alg:standard_tensor_algorithm}.

\begin{algorithm}
\caption{Algorithm for calculating the tensor corresponding to the standard matrix multiplication algorithm}
\label{alg:standard_tensor_algorithm}
\hspace*{\algorithmicindent} \textbf{Input:} For multiplying $n \times m$ and $m \times p$ matrices, input $n$, $m$, $p$ \\
\hspace*{\algorithmicindent} \textbf{Output:} Tensor encoding standard matrix multiplication having dimension $(n \times m, m \times p, p \times n)$

\begin{algorithmic}[1]
\State $tensor \gets \text{zero tensor of size } (n \times m) \times (m \times p) \times (p \times n)$
\For{$i \gets 0$ \textbf{to} $n-1$}
    \For{$j \gets 0$ \textbf{to} $m-1$}
        \For{$k \gets 0$ \textbf{to} $p-1$}
            \State $tensor[i \times m + j][j \times p + k][k \times n + i] \gets 1$
        \EndFor
    \EndFor
\EndFor
\State \Return $tensor$
\end{algorithmic}
\end{algorithm} 

\begin{figure*}[tbp]
    \centering
    \includegraphics[width=0.7\textwidth]{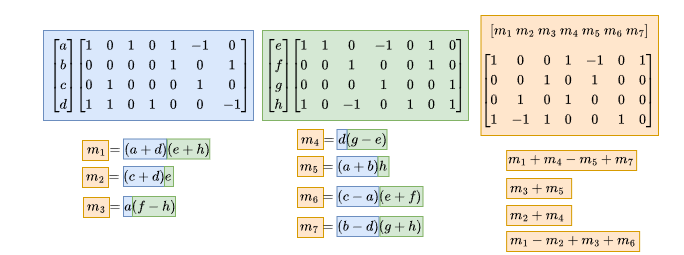}
    \caption{Tensor decomposition for the Strassen algorithm. Every column in the blue and green matrices gives us the coefficients for the elements that should be multiplied. For example, the first columns are $[1,0,0,1]$, which corresponds to sums $a + d$ (blue) and $e + h$ (green). The $m_i$ values are summed over the rows of the third matrix. For example, row $[1,0,0,1-1,0,1]$ corresponds to sum $m_1 + m_4 - m_5 + m_7$. These values work as an input for Alg.~\ref{alg:meta_algorithm}.}
    \label{fig:Strassens_algorithm}
\end{figure*}

To tackle the problem of finding tensor decompositions, we proceed in the following way. In practice, we search for vectors $\mathbf{x}_i \in \mathds{R}^{n \times m}$, $\mathbf{y}_i \in \mathds{R}^{m \times p}$ and $\mathbf{z}_i \in \mathds{R}^{n \times p}$ so that
\begin{equation}\label{eq:tensor_decompostion_xyz}
    M_{n,m,p} = \sum_{i = 1}^{R} \mathbf{x}_i \otimes \mathbf{y}_i \otimes \mathbf{z}_i.
\end{equation}
For example, Fig.~\ref{fig:Strassens_algorithm} shows that the first vectors in the decomposition of Strassen's algorithm are $x_1 = y_1 = z_1 = [1,0,0,1]^{\top}$, which correspond to the first columns of the matrices in Fig.~\ref{fig:Strassens_algorithm}. The second column provides the vectors $x_2$, $y_2$, $z_2$, etc. From the tensor decompositions, we obtain algorithms to multiply matrices. The following Alg. \ref{alg:meta_algorithm} is the generalized version of the algorithm presented in \cite{Fawzi_Balog_Huang_Hubert_Romera_Paredes_Barekatain_Novikov_RRuiz_Schrittwieser_Swirszcz_et_2022}, and it describes how the abstract tensor decompositions \eqref{eq:tensor_decompostion_xyz} are mapped back to practical matrix multiplication algorithms. The algorithm generalizes the idea we represented in subsection \ref{subsection:strasse} for the Strassen algorithm.

\begin{algorithm}
\caption{Meta-algorithm for transforming a tensor decomposition into a matrix multiplication algorithm}
\label{alg:meta_algorithm}

\begin{algorithmic}[1]
\State \textbf{Parameters}: $\{\mathbf{x}_i, \mathbf{y}_i, \mathbf{z}_i\}_{i=1}^R$ vectors such that $T_{n,m,p} = \sum_{i=1}^R \mathbf{x}_i \otimes \mathbf{y}_i \otimes \mathbf{z}_i$
\State \textbf{Input}: matrix $A$ of size $n \times m$, matrix $B$ of size $m \times p$
\State \textbf{Output}: matrix multiplication result $C = AB$
\For{$i = 1$ \textbf{to} $R$}
    \State $m_i \gets (a_1 x_1^{i} + \ldots + a_{nm}x_{nm}^{i}) \cdot (b_1y^{i}_{1} + \ldots + b_{mp}y^{i}_{mp})$
\EndFor
\For{$i = 1$ \textbf{to} $np$}
    \State $c_i \gets m_1 z_{1}^{i} + \ldots + m_R z_{R}^{i}$
\EndFor
\State \Return $C$
\end{algorithmic}
\end{algorithm}

In this work, we consider two cases depending on the scalar field. We assume that the multiplied matrices have terms in $\mathds{R}$ or in $\mathds{F}_{2} = \left\{0,1 \right\}$ which is the finite two-element field where the addition is defined by addition modulo two and multiplication follows the multiplication of $0$ and $1$ in reals. A similar division was also made in \cite{Fawzi_Balog_Huang_Hubert_Romera_Paredes_Barekatain_Novikov_RRuiz_Schrittwieser_Swirszcz_et_2022}.

\section{Algorithms}\label{sec:algorithms}

We start this section by describing the main objective that the decompositional and holistic algorithms share. The main objective utilizes Hamming distance for tensors. Since we are focused on multiplying two matrices, we give the Hamming distance definition only for three-dimensional tensors. For three-dimensional tensors $T_1$ and $T_2$ of both having dimension $(n, m, p)$, the Hamming distance is defined as
\begin{equation*}
    d_H(T_1, T_2) := \sum_{i_1 = 1}^{n}\sum_{i_2 = 1}^{m}\sum_{i_3 = 1}^{p} |T_1(i_1, i_2, i_3) - T_2(i_1, i_2, i_3)|,
\end{equation*}
where the absolute value of the difference is calculated over the scalar values in the tensors. As we noted before, we focus on scalar fields $\mathds{R}$  and  $\mathds{F}_2$. Now, we state the main objective whose variations the decompositional and holistic algorithms aim to minimize.

\begin{center}
\begin{tcolorbox}[width=.95\columnwidth, standard jigsaw, title={Main objective}, opacityback=0, label = prob:main_objective]
    Let $S$ be a (source) tensor and $T$ be a (target) tensor, both having dimension $(n, m, p)$. Select vectors $\mathbf{x}$, $\mathbf{y}$ and $\mathbf{z}$ having dimensions $n$, $m$, and $p$, respectively, so that the Hamming distance
    \begin{equation*}
        d_H(T, S - \mathbf{x} \otimes \mathbf{y} \otimes \mathbf{z})
    \end{equation*}
    is minimized.
\end{tcolorbox}
\end{center}

An intuitive way to understand this is to consider that we aim to move source tensor $S$ towards target tensor $T$ with respect to the Hamming distance by subtracting the tensor product of vectors $\mathbf{x}$, $\mathbf{y}$ and $\mathbf{z}$. Now we let a quantum computer or classical optimizer find vectors $\mathbf{x}$, $\mathbf{y}$ and $\mathbf{z}$ that minimize this distance. Thus, we define that the terms in vectors $\mathbf{x}$, $\mathbf{y}$, and $\mathbf{z}$ are the variables we are optimizing. Generally, they are either integer or binary variables.

The standard fact of metrics implies that $d_H(T, S - \mathbf{x} \otimes \mathbf{y} \otimes \mathbf{z}) = 0$ if and only if $T = S - \mathbf{x} \otimes \mathbf{y} \otimes \mathbf{z}$ and otherwise the distance is positive. The reason to use Hamming distance is that it produces HUBO problems naturally. By using the definition of Hamming distance, we obtain
\begin{align}\label{eq:hubo}
    d_H&(T, S - \mathbf{x} \otimes \mathbf{y} \otimes \mathbf{z}) \notag \\ &= \sum_{i = 1}^{nm}\sum_{j = 1}^{mp}\sum_{k = 1}^{pn} |T(i, j, k) - S(i, j, k) + x_{i}y_{j}z_{k}|.
\end{align}

Formulation \eqref{eq:hubo} is not a HUBO problem yet. We have to ensure that the variables are binary and that we can write the objective without the absolute value. 

First, we focus on the problem of encoding the variables. If we are optimizing over the field $\mathds{F}_2$, we need to consider only binary variables. More precisely, we state that $\mathbf{x} = (x_1, \ldots, x_{nm})$, $\mathbf{y} = (y_1, \ldots, y_{mp})$ and $\mathbf{z} = (z_1, \ldots, z_{pn})$ are binary variable vectors and we do not need to modify them. In the case that the tensor decomposition is computed over $\mathds{R}$, we use integer variables. As noted in \cite{Fawzi_Balog_Huang_Hubert_Romera_Paredes_Barekatain_Novikov_RRuiz_Schrittwieser_Swirszcz_et_2022}, it suffices to consider integer variables that range over a relatively small set of integers such as $\left\{ -2, -1, 0, 1, 2 \right\}$. We then can employ the standard technique from \cite{lucas_2014} to rewrite integer variables using binary variables. Assume $x$ is a bounded integer variable that can have $N$ distinct values such that $2^M \leq N < 2^{M + 1}$, then
\begin{equation*}
    x = \sum_{n = 1}^{N}nx_n \rightarrow \sum_{n = 0}^{M-1} 2^nx_n + (N + 1 - 2^{M})x_M,
\end{equation*}
where $x_n$ for $n = 1, \ldots, N$ are binary variables.

Next, we discuss how to write \eqref{eq:hubo} without the absolute values. Depending on the terms in the tensors and the scalar field, we have different ways to rewrite \eqref{eq:hubo} without absolute values. The standard technique would be to square the terms
\begin{align*}
    \sum_{i = 1}^{nm}\sum_{j = 1}^{mp}\sum_{k = 1}^{pn} (T(i, j, k) - S(i, j, k) + x_{i}y_{j}z_{k})^2.
\end{align*}
This approach is relatively costly since it introduces more higher-order terms to the polynomial if we are using previous techniques to encode integer variables. On the other hand, it keeps the minimum of the objective function unchanged.

If the elements $T(i, j, k)$ and $S(i, j, k)$ are in field $\mathds{F}_2$, we can rewrite \eqref{eq:hubo} without the absolute value so that
\begin{align*}
    |T(i, j, k)& - S(i, j, k) + x_{i}y_{j}z_{k}| \\ &= \begin{cases}
    1 - x_{i}y_{j}z_{k}, &\text{ if } T(i, j, k) - S(i, j, k) = 1 \\
    x_{i}y_{j}z_{k} &\text{ otherwise.}
    \end{cases}
\end{align*}
We can see that the previous rewriting actually encodes the original problem. If $T(i, j, k) - S(i, j, k) = 1$, then $|T(i, j, k) - S(i, j, k) + x_{i}y_{j}z_{k}|$ is $0$ when $x_{i}y_{j}z_{k} = 1$ and $1$ when $x_{i}y_{j}z_{k} = 0$. Similarly, $1 - x_{i}y_{j}z_{k}$ is $0$ when $x_{i}y_{j}z_{k} = 1$ and $1$ when $x_{i}y_{j}z_{k} = 0$. The case that $T(i, j, k) - S(i, j, k) = 0$ is similar.

In the case that we restrict the variable vectors to consist of binary variables, \eqref{eq:hubo} defines a HUBO problem since we have the cubic terms $x_{i}y_{j}z_{k}$. In all of the previously described cases, we obtain a HUBO problem. By minimizing the distance, i.e., minimizing the corresponding HUBO problem, we obtain a configuration for the binary variables so that the source tensor is moved closer to the target tensor with respect to the Hamming distance.

In conclusion, we described how to perform the optimization for a single triple of vectors $\mathbf{x}$, $\mathbf{y}$, and $\mathbf{z}$. In actual tensor decomposition calculations, we need to find a sequence $(\mathbf{x}_i, \mathbf{y}_i, \mathbf{z}_i)_{i = 1}^{R}$ of vectors so that the given tensor is decomposed. The decompositional and holistic algorithms answer the question of how to perform this overall optimization process. Finally, we summarize a single-step optimization in Alg.~\ref{alg:move_tensors_closer}. 

\begin{algorithm}
\caption{Move Tensors Closer Subroutine}
\label{alg:move_tensors_closer}
\begin{algorithmic}[1]
\Procedure{MoveTensorsCloser}{$S$, $T_s$, field}
    \State $\mathbf{x}$, $\mathbf{y}$, $\mathbf{z}$ $\gets$ initialize\_variables(field)
    \State HUBO $\gets$ $d_H(S, T_s - \mathbf{x} \otimes \mathbf{y} \otimes \mathbf{z})$
    \State QUBO $\gets$ make\_quadratic(HUBO)
    \State sample $\gets$ solve(QUBO)
    \State x, y, z $\gets$ construct\_vectors\_from\_sample(sample)
    \State \Return $x \otimes y \otimes z$
\EndProcedure
\end{algorithmic}
\end{algorithm}

\subsection{Decompositional algorithm}

The key difference between the decompositional and holistic algorithms is that in the decompositional algorithm, we decompose the problem into sub-objectives, whereas in the holistic algorithm, we aim to solve all the steps at once. Due to the decompositionality of this approach, we have feasible HUBO and QUBO problems, whereas the holistic approach produces objectives whose optimal was not found with the current quantum or classical devices.

Let $\mathcal{O}$ be the zero tensor and $T_s$ be the tensor encoding the standard matrix multiplication. Tensor $T_s$ can be computed with Alog.~\ref{alg:standard_tensor_algorithm}. More formally, the difference between decompositional and holistic algorithms is in the summation of the tensors, which are tensor products of rank-one tensors. In the case of the decompositional algorithm, we minimize first and then sum. Thus, we obtain the following total objective:
\begin{equation}
    \mathrm{argmin} \quad  \sum_{i = 1}^{R} d_H(\mathcal{O}, T_i - \mathbf{x}_i \otimes \mathbf{y}_i \otimes \mathbf{z}_i),
\end{equation}
where $T_i = T_{i-1} - \mathbf{x}_i \otimes \mathbf{y}_i \otimes \mathbf{z}_i$ for $i = 2, \ldots, R$ and $T_1 = T_s$. The minimization is performed with respect to the binary variables in vectors $\mathbf{x}_i$, $\mathbf{y}_i$, and $\mathbf{z}_i$ for $i = 1, \ldots, R$.

In the holistic approach, we sum first and then minimize. Thus, we have the following minimization problem
\begin{equation*}
    \mathrm{argmin} \quad d_H(\mathcal{O}, T_s - \sum_{i = 1}^{R} \mathbf{x}_i \otimes \mathbf{y}_i \otimes \mathbf{z}_i),
\end{equation*}
which we will describe in more detail later.

The first but sub-optimal idea is to find a sequence of vectors $\mathbf{x}_i$, $\mathbf{y}_i$, and $\mathbf{z}_i$ so that we move from standard tensor $T_s$ to origo $\mathcal{O}$ by minimizing the Hamming distance in the main objective at every step. While this method works and produces a tensor decomposition, the main objective does not guarantee that the decomposition has a minimal total rank. In other words, we often obtain trivial decompositions, such as the standard matrix multiplication, or even longer decompositions. For example, in the case of $2 \times 2$ matrices, we discover the standard matrix multiplication algorithm instead of the Strassen algorithm.

To improve the probability of finding faster matrix multiplication algorithms, we modify the decompositional algorithm to start the optimization from a so-called high-energy point. The core idea is that we select an initial point $T_{\text{high}}$ and also fix an initial tensor $\mathbf{x}_{\text{high}} \otimes \mathbf{y}_{\text{high}} \otimes \mathbf{z}_{\text{high}}$ that we use to start the minimization process. We aim to select these points so that minimizing the Hamming distance at every step would minimize the total rank. As we will demonstrate with the Strassen algorithm, this means that the Hamming distance between tensor $T_{\text{high}}$ and $\mathcal{O}$ should be large. That is why we call the tensor $T_{\text{high}}$ a high-energy tensor. 

\begin{figure}[tbp]
    \centering
    \includegraphics[width = .99\columnwidth]{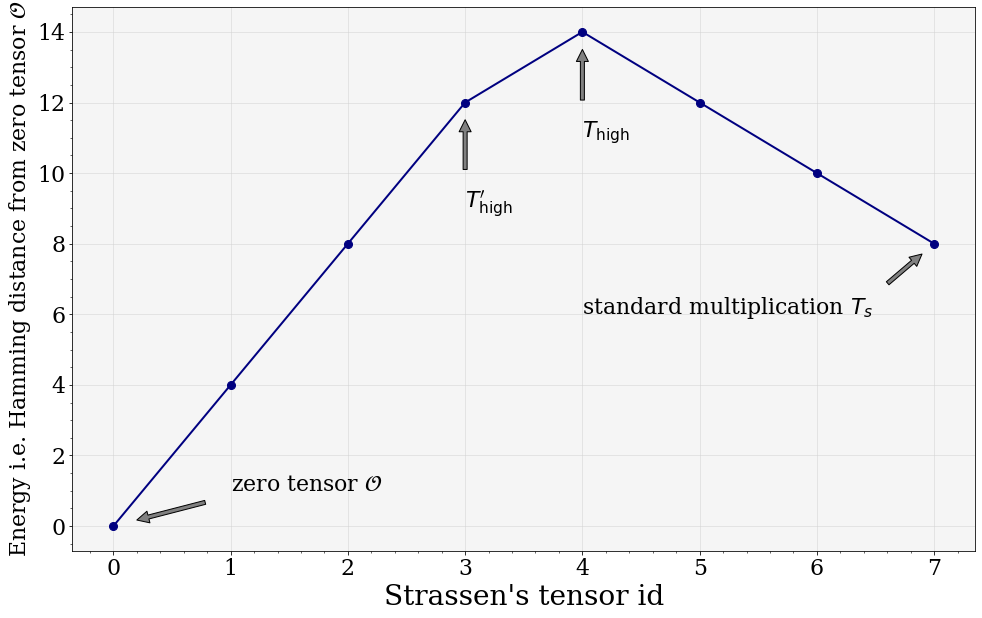}
    \caption{Energy landscape of the optimization problem to discover the Strassen matrix multiplication algorithm}
    \label{fig:strasse_energy_landscape}
\end{figure}

In order to utilize the good initial point $T_{\text{high}}$, we divide the decompositional optimization into two parts. The first part starts from the high-energy tensor $T_{\text{high}}$. It aims to advance towards the standard matrix multiplication tensor $T_s$ by minimizing the Hamming distance as we described at the beginning of this section. The second part starts from the tensor $T_{\text{high}}' = T_{\text{high}} - \mathbf{x}_{\text{high}} \otimes \mathbf{y}_{\text{high}} \otimes \mathbf{z}_{\text{high}}$ and aims to advance towards zero tensor $\mathcal{O}$. Thus, from the first optimization part, we obtain the decomposition $T_{\text{high}} = T_s - \sum_{i = 1}^{R_0} \mathbf{x}_{i} \otimes \mathbf{y}_{i} \otimes \mathbf{z}_{i}$. The second part produces $T_{\text{high}}' = \mathcal{O} - \sum_{i = 1}^{R_1} \mathbf{x}_{i} \otimes \mathbf{y}_{i} \otimes \mathbf{z}_{i}$. Altogether, with slightly modified indexing, we obtain
\begin{equation*}
    T_s = \mathbf{x}_{\text{high}} \otimes \mathbf{y}_{\text{high}} \otimes \mathbf{z}_{\text{high}} + \sum_{i = 1}^{R_0 + R_1} \mathbf{x}_{i} \otimes \mathbf{y}_{i} \otimes \mathbf{z}_{i},
\end{equation*}
which is rank $R_0 + R_1 + 1$ decomposition of the standard matrix multiplication algorithm. Fig.~\ref{fig:strasse_energy_landscape} shows the real values for the tensor decomposition corresponding to the Strassen matrix multiplication algorithm. By starting from the high energy points $T_{\text{high}}$ and $T_{\text{high}}'$ and ''sliding downhill'' towards $\mathcal{O}$ and $T_s$, we are able to find the Strassen matrix multiplication algorithm in the case of $2 \times 2$ matrices. On the other hand, in the other cases, we noticed that the performance of the algorithm highly depends on the initial values $T_{\text{high}}$ and $T_{\text{high}}'$. 

We summarize the full decompositional algorithm with an initial point in Algorithm \ref{alg:modular_algorithm}. The algorithm relies on the MoveTensorsCloser subroutine described in Alg.~\ref{alg:move_tensors_closer}.

\begin{algorithm}
\caption{Full Modular Algorithm}
\label{alg:modular_algorithm}
\begin{algorithmic}[1]
\State \textbf{Input}: $T_s$, $\mathcal{O}$, $T_{\text{high}}$, $\mathbf{x}_{\text{high}} \otimes \mathbf{y}_{\text{high}} \otimes \mathbf{z}_{\text{high}}$, field
\State tensor $\gets$ $T_{\text{high}}$
\State decomposition $\gets$ [$\mathbf{x}_{\text{high}} \otimes \mathbf{y}_{\text{high}} \otimes \mathbf{z}_{\text{high}}$]
\Loop
    \State $x \otimes y \otimes z$ $\gets$ \Call{MoveTensorsCloser}{$T_s$, tensor, field}
    \State append $x \otimes y \otimes z$ to decompositions
    \If{field = $\mathds{R}$}
        \State tensor $\gets$ tensor $- x \otimes y \otimes z$
    \ElsIf{field = $\mathds{F}_2$}
        \State tensor $\gets$ tensor $- x \otimes y \otimes z \ \mathrm{mod} \ 2$
    \EndIf
    \If{tensor = $T_s$}
        \State \textbf{break}
    \EndIf
\EndLoop
\State tensor $\gets$ $T_{\text{high}} - \mathbf{x}_{\text{high}} \otimes \mathbf{y}_{\text{high}} \otimes \mathbf{z}_{\text{high}}$
\Loop
    \State $x \otimes y \otimes z$ $\gets$ \Call{MoveTensorsCloser}{$\mathcal{O}$, tensor, field}
    \State append $x \otimes y \otimes z$ to decompositions
    \If{field = $\mathds{R}$}
        \State tensor $\gets$ tensor $- x \otimes y \otimes z$
    \ElsIf{field = $\mathds{F}_2$}
        \State tensor $\gets$ tensor $- x \otimes y \otimes z \ \mathrm{mod} \ 2$
    \EndIf
    \If{tensor = $\mathcal{O}$}
        \State \textbf{break}
    \EndIf
\EndLoop
\State \Return decomposition
\end{algorithmic}
\end{algorithm}

\textbf{Computational complexity of the decompositional algorithm.} Every application of Algorithm \ref{alg:move_tensors_closer} requires creating a HUBO problem that consists of $nm + mp + pn \leq 3\max(n,m,p)^2$ variables. Depending if we are dealing with binary variables or integer variables, the final number of variables varies. In the case of binary variables, we do not need to increase the number. On the other hand, when we are encoding integer variables into binary variables, we need to introduce $\lceil \log(M) \rceil$ new variables \cite{lucas_2014} where $2^M \leq N < 2^{M +1}$ and $N$ is the number of integers we need to encode. 

Although the number of variables stays reasonable for the matrices considered in \cite{Fawzi_Balog_Huang_Hubert_Romera_Paredes_Barekatain_Novikov_RRuiz_Schrittwieser_Swirszcz_et_2022}, the challenge rises from the number of interactions. Creating the tensor product of variable vectors of sizes $nm$, $mp$, and $pn$ produces a number of $n^2m^2p^2 \leq \max(n,m,p)^6$ cubic interactions. Furthermore, to reduce the cubic interactions into quadratic interactions, we need to introduce at most $\max(n,m,p)^6$ ancillary variables \cite{boros_2002}. Since the cases we have been considering allow us to assume $n \leq m \leq p$, we summarize the qubit complexity as $\mathcal{O}(p^2)$ and the interaction complexity as $\mathcal{O}(p^6)$.

The decompositional algorithm described in Algorithm \ref{alg:modular_algorithm} consists of two loops that can be executed in parallel utilizing two quantum computers simultaneously. If we are interested in discovering faster matrix multiplication algorithms, we need to execute each of the loops only $\lceil (R - 1)/2 \rceil$ times where $R$ is the best-known rank for the multiplication.

\subsection{Holistic algorithm for fixed length tensor decompositions}

Unconstrained binary optimization problems often minimize with respect to a fixed length that the solution should obey \cite{lucas_2014}. In the holistic approach, we fix the length, i.e., the rank of the tensor decomposition, and aim to optimize the whole decomposition in a single step. Let the fixed rank be $R$. Fixing a rank that is less than the best-known for multiplying $(n,m,p)$ matrices might produce a better decomposition for the multiplication algorithm if a solution exists. If a solution does not exist, the minimum of the objective function, i.e., the polynomial encoding HUBO problem, is positive.

As noted at the beginning of the previous subsection, the minimization problem in the holistic algorithm is
\begin{equation}\label{eq:holistic_objective}
    \mathrm{argmin} \quad d_H(\mathcal{O}, T_s - \sum_{r = 1}^{R} \mathbf{x}_r \otimes \mathbf{y}_r \otimes \mathbf{z}_r),
\end{equation}
where we aim to minimize with respect to the variable tensors $\mathbf{x}_r \otimes \mathbf{y}_r \otimes \mathbf{z}_r$ for $r = 1, \ldots, R$. When we use the definition of Hamming distance and simplify the expression, we obtain
\begin{equation}\label{eq:holistic_opened}
    \sum_{i = 1}^{nm}\sum_{j = 1}^{mp}\sum_{k = 1}^{pn} (T_s(i, j, k) -  \sum_{r = 1}^{R} x_{i}^r y_{j}^r z_{k}^r)^2.
\end{equation}

This formulation defines a HUBO problem that can be converted into QUBO with the rewriting schemes described in \ref{subsection:hubo}.

If $x_1, \ldots, x_n$ are binary variables, the QUBO objective function is an element of the polynomial field $\mathds{R}[x_1, \ldots, x_n]$. Since quantum annealers and, as far as we know, all classical solvers aim to optimize with respect to this field. We are not able to natively encode problems that should be minimized over the polynomial field $\mathds{F}_2[x_1, \ldots, x_n]$. We were lucky in the decompositional approach that minimizing the objective function in $\mathds{R}[x_1, \ldots, x_n]$ matched with minimizing it in $\mathds{F}_2[x_1, \ldots, x_n]$. This is not the case in the holistic algorithm. 

In the implementation, we have especially focused on encoding integer set $\left\{ -1, 0, 1 \right\}$. This can be achieved by replacing every integer variable $x$ with two new binary variables $x_{left} - x_{right}$. If $x_{left} = x_{right}$, then $x = 0$. If $x_{left} = 1$ and $x_{right} = 0$, then $x = 1$. If $x_{left} = 0$ and $x_{right} = 1$, then $x = -1$. Thus, we can encode the integer variables $x \in\left\{ -1, 0, 1 \right\}$ using the binary variables  $x_{left}$ and $x_{right}$.

The pseudocode for the holistic algorithm is almost identical to Algorithm \ref{alg:move_tensors_closer} except that we change line three to include the objective from \eqref{eq:holistic_objective} and we execute the algorithm only once. If we are able to find such a binary variable configuration that the objective in \eqref{eq:holistic_objective} reaches zero, we have found a valid rank $R$ decomposition for tensor $T_s$. Otherwise, the value of the objective function is positive, and the decomposition is not valid.

\textbf{Computational complexity.} In the case of the decompositional algorithm, we concluded that we obtain $nm + mp + pn$ variables. In the holistic algorithm, we have $R$ number of these variables leading to a total $R(nm + mp + pn)$ variables. Assuming that $n \leq m \leq p$, the number of required qubits is $\mathcal{O}(Rp^2)$. Furthermore, we are dealing with integer variables leading to the similar requirement of introducing $\lceil \log(M) \rceil$ ancillas that encode the integer variables. 

The number of quadratic or higher degree interactions is large for the holistic algorithm in practice. Let us consider that we use $k$ binary variables to encode the integer variables in \eqref{eq:holistic_opened}. Using $k$ binary variables, we obtain $k^3$ cubic interactions for each $xyz$ integer interaction. In total, we have a number of $R$ cubic integer interactions, which leads to $Rk^3$ cubic interactions between binary variables. In the worst case, target tensor $T_s$ contains only non-zero elements. Thus, we can obtain $(Rk^3 + 1)^2$ cubic interactions between binary variables. Finally, in total, we have $n^2m^2p^2$ of the elements consisting of the cubic interactions, which leads to $n^2m^2p^2(Rk^3 + 1)^2$ interactions. 

\begin{figure}[tbp]
    \centering
    \includegraphics[width = .99\columnwidth]{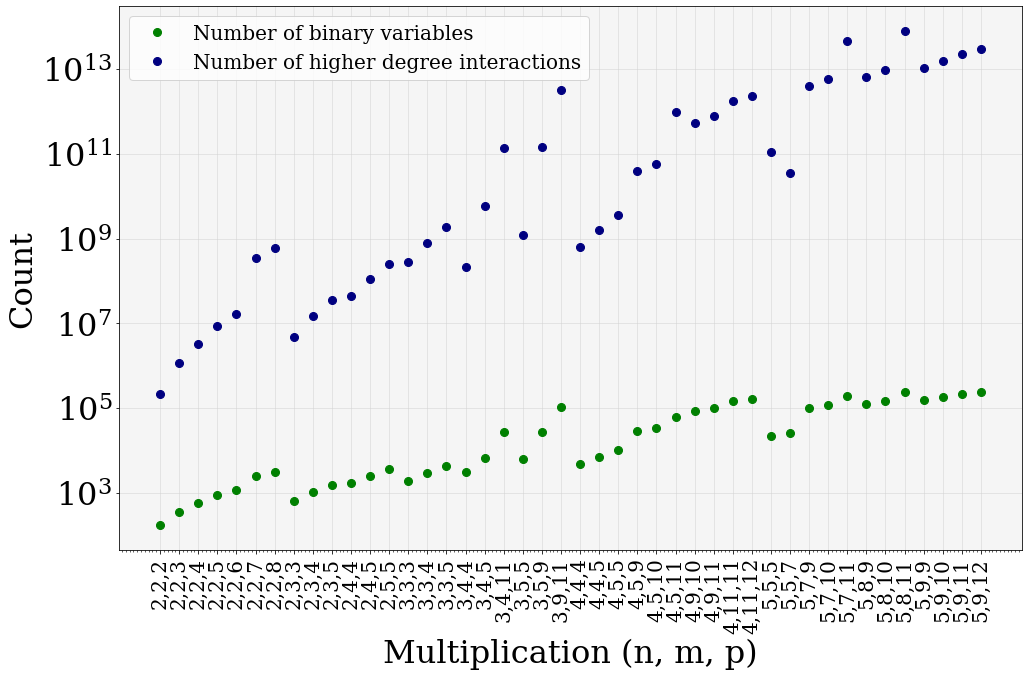}
    \caption{Worst case scalability of the holistic algorithm}
    \label{fig:holistic_scalability_1}
\end{figure}

We have demonstrated the worst-case scalability of this algorithm in Fig.~\ref{fig:holistic_scalability_1} varying sizes of the multiplied matrices. To summarize the figure, we see that the number of binary variables stays below a million in all of the plotted cases. A million variables is a lot for the current universal quantum computers, but probably not impossible for the near-future devices. Although the scalability shows challenges for the current devices, it also shows the problem's property that the number of variables is relatively small compared to the number of interactions (over $10^{13}$ interactions). 

\section{Implementation and results}

In this section, we demonstrate the applicability of the proposed algorithms in practice. For reproducibility, the implementation and the HUBOs can be found on GitHub \cite{uotila_2023}.

The technical implementation is utilizing D-wave's Ocean software framework \cite{ocean}. We have also integrated classical solvers, IBM's CPLEX \cite{cplex}, and Gurobi \cite{gurobi}, to accept and minimize the QUBO problems. The classical solvers are primarily designed for linear programming, mixed-integer linear programming, and quadratic programming. D-wave and the classical solvers are not able to solve HUBO problems without rewriting them as QUBO problems.

\subsection{Discover Strassen algorithm}

In this subsection, we will explain how the Strassen algorithm can be discovered using the decompositional algorithm. We continue the theory we developed earlier in \ref{subsection:strasse}. We will also point out an interesting symmetry property in the tensor decomposition which has a connection to the energy values of the Hamiltonians in the decompositional algorithm. 

Next, we explain how we discover the Strassen algorithm with the decompositional algorithm in the case that the scalar field is $\mathds{F}_2$. The binary variables in the HUBO and QUBO problems are the elements in the following binary vectors:
\begin{align*}
    \textbf{x}_i = [x^{i}_1, x^{i}_2, &x^{i}_3, x^{i}_4]^{\top}, \ \textbf{y}_i = [y^{i}_1, y^{i}_2, y^{i}_3, y^{i}_4]^{\top}, \\ &\textbf{z}_i = [z^{i}_1, z^{i}_2, z^{i}_3, z^{i}_4]^{\top},
\end{align*}
for $i = 1, \ldots, 7$. We want to use seven vectors because the naive matrix multiplication algorithm uses eight vectors, and we want to find a faster algorithm.

Next, we describe how we can find a good initial high-energy tensor $T_{\text{high}}$ and the corresponding vectors $\mathbf{x}_{\text{high}}$, $\mathbf{y}_{\text{high}}$, $\mathbf{z}_{\text{high}}$ for the decompositional algorithm. To this end, it is important to highlight some characteristics of the tensors involved in the decomposition of the Strassen algorithm. By definition, three vectors represent each tensor in the decomposition. As we aggregate these vectors into matrices at each step, we derive the following matrices where each row corresponds to a vector.

\begin{align}\label{eq:symmetries}
&\begin{bmatrix} 1 & 1 & 0 & 0 \\ 0 & 0 & 0 & 1 \\ -1 & 1 & 0 & 0 \end{bmatrix}, &\begin{bmatrix} 0 & 0 & 1 & 1 \\ 1 & 0 & 0 & 0 \\ 0 & 0 & 1 & -1 \end{bmatrix}, \notag \\
&\begin{bmatrix} 0 & 0 & 0 & 1 \\ -1 & 0 & 1 & 0 \\ 1 & 0 & 1 & 0 \end{bmatrix}, &\begin{bmatrix} 1 & 0 & 0 & 0 \\ 0 & 1 & 0 & -1 \\ 0 & 1 & 0 & 1 \end{bmatrix}, \\
&\begin{bmatrix} -1 & 0 & 1 & 0 \\ 1 & 1 & 0 & 0 \\ 0 & 0 & 0 & 1 \end{bmatrix}, &\begin{bmatrix} 0 & 1 & 0 & -1 \\ 0 & 0 & 1 & 1 \\ 1 & 0 & 0 & 0 \end{bmatrix}, \notag
\end{align}
 \begin{displaymath}
     \begin{bmatrix} 1 & 0 & 0 & 1 \\ 1 & 0 & 0 & 1 \\ 1 & 0 & 0 & 1 \end{bmatrix}
 \end{displaymath}
 
The matrices can also be seen in Fig.~\ref{fig:Strassens_algorithm}, where they represent the columns of the three matrices (blue, green, red). Matrices are solutions to the QUBO problems solved at each step in the decompositional algorithm. Moreover, we have grouped the matrices in a special way, which should reveal the symmetries between them: at each row in \eqref{eq:symmetries}, the left matrix is symmetric to the right matrix. Interestingly, AlphaTensor did not rediscover exactly the same tensor decomposition for the Strassen algorithm, which we find with our method. Now, we compute tensor $T_{\text{high}}$ so that it is constructed based on the matrices in the first column in \eqref{eq:symmetries}. Tensor $T_{\text{high}}$ is the suitable initial point for the decompositional algorithm. Using the previous notation, this means that we compute
\begin{align*}
    T_{\text{high}} &= T_s - [1, 1, 0, 0]^{\top} \otimes [0, 0, 0, 1]^{\top} \otimes [-1, 0, 1, 0]^{\top} \\
&- [0, 0, 0, 1]^{\top} \otimes [-1, 0, 1, 0]^{\top} \otimes [1, 0, 1, 0]^{\top} \\
&- [-1, 0, 1, 0]^{\top} \otimes [1, 1, 0, 0]^{\top} \otimes [0, 0, 0, 1]^{\top},
\end{align*}
where each tensor is computed using the rows of the matrices in the first column in \eqref{eq:symmetries}. Moreover, we compute the point
\begin{displaymath}
    \mathbf{x}_{\text{high}} = [1, 0, 0, 1]^{\top}, \ \mathbf{y}_{\text{high}} = [1, 0, 0, 1]^{\top}, \ \mathbf{z}_{\text{high}} = [1, 0, 0, 1]^{\top},
\end{displaymath}
which is formed based on rows of the matrix that is the last in \eqref{eq:symmetries}. This selection also respects the symmetry that appears between the matrices in \ref{eq:symmetries}. Note that although we obtained $T_{\text{high}}$ in a way that relies on the known decomposition for the Strassen algorithm, the tensor $T_{\text{high}}$ does not encode any information about the decomposition and it does not make the decomposition problem easier.

Now that we have $T_{\text{high}}$ and $\mathbf{x}_{\text{high}}$, $\mathbf{y}_{\text{high}}$, and $\mathbf{z}_{\text{high}}$, we compute the tensor decomposition for the Strassen algorithm using the decompositional algorithm described in Alg. \ref{alg:modular_algorithm}. At each step, the minimum to the QUBO problem provides the optimal vectors $\textbf{x}_i$, $\textbf{y}_i$, and $\textbf{z}_i$, which are exactly (modulo 2) the rows of the first six matrices in \eqref{eq:symmetries}. The algorithm can be executed with the code provided on GitHub \cite{uotila_2023}. The steps are also visible in Fig.~\ref{fig:strasse_energy_landscape} where the decomposition algorithm starts from the tensors $T_{\text{high}}$ and $T'_{\text{high}} = T_{\text{high}} - \mathbf{x}_{\text{high}} \otimes \mathbf{y}_{\text{high}} \otimes \mathbf{z}_{\text{high}}$ and ''slides downhill'' towards the standard multiplication tensor and the zero tensor.

\subsection{Evaluation of holistic algorithm}

In this subsection, we focus on evaluating the holistic algorithm for finding decompositions for multiplying $2 \times 2$ matrices. As noted before, the holistic algorithm works only in the field $\mathds{R}$, and thus, we have to perform the relatively costly integer variable to binary variable transformation. We have constructed the HUBO objective for multiplying $2 \times 2$ matrices defined in \eqref{eq:holistic_objective}.

\begin{figure}[tbp]
    \centering
    \includegraphics[width = .99\columnwidth]{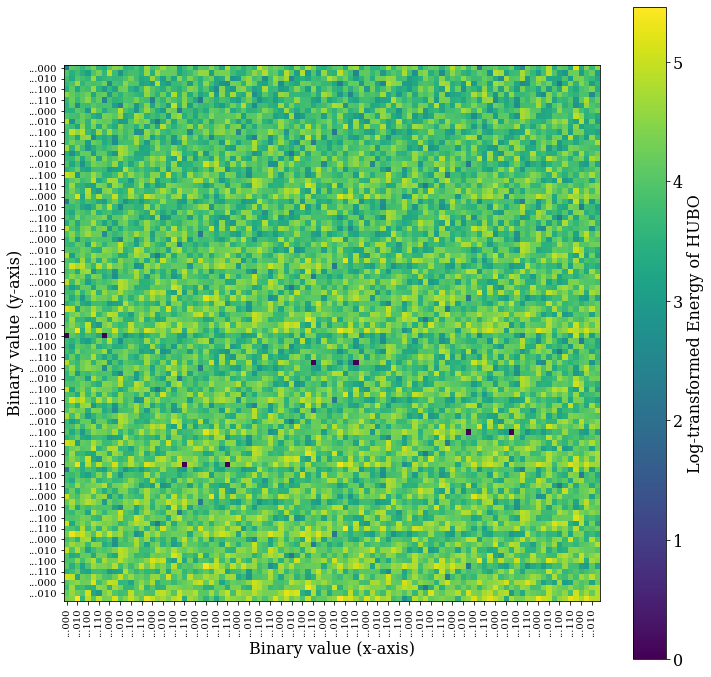}
    \caption{Overall view to the optimization landscape in the holistic algorithm in the case $2,2,2$. The eight optimal binary variable configurations are visible in the space.}
    \label{fig:holistic_overview}
\end{figure}

Fig.~\ref{fig:holistic_overview} demonstrates the energy landscape for this HUBO problem. As discussed in the computational complexity part of the holistic algorithm, in this case, we have $168$ binary variables, which result in $2^{168}$ different binary value combinations. These binary value combinations are necessarily binary numbers that have the natural order. We divide this binary number interval (representing $[0, 2^{168}]$) into equal-sized chunks and calculate the value of the HUBO objective at each chunk at a randomly selected point. Plotting the energy value at the selected points, we obtain Fig.~\ref{fig:holistic_overview}. We intentionally insert the known optimal values into the space to their correct positions to demonstrate that the HUBO correctly reaches $0$ in the valid configurations being positive elsewhere. The valid configurations, i.e., the optimal solutions to the problem, correspond to the Strassen algorithm. There are multiple valid configurations due to transforming integer variables into binary variables, but all these combinations produce the same matrix multiplication algorithm. Moreover, we have log-transformed the energy values to make the optimal points visible.

\begin{figure}[tbp]
    \centering
    \includegraphics[width = .99\columnwidth]{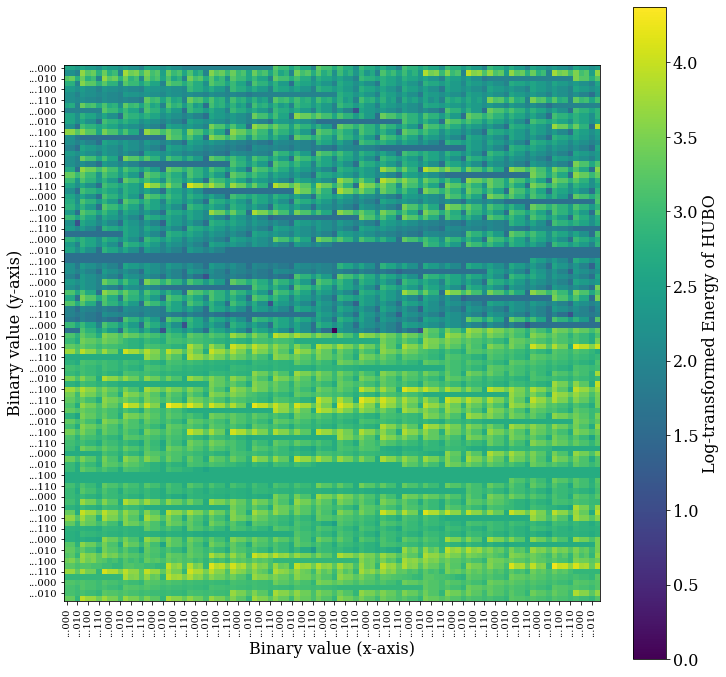}
    \caption{Visualization of the optimization landscape when we have zoomed in the neighborhood of one of the optimal points for the HUBO objective. The optimal solution is visible in the middle.}
    \label{fig:strassen_heatmap_zoomed}
\end{figure}

Interestingly, the first solution is only approximately $0.00152\%$ off from the midpoint $2^{167}$ of the interval $[0, 2^{168}]$. This might indicate that the search process for finding the optimal point for multiplying $n \times n$ matrices should be started around the midpoint of the search interval. In Fig.~\ref{fig:strassen_heatmap_zoomed}, we have zoomed into the neighborhood of the first optimal point in Fig~\ref{fig:holistic_overview}. We have plotted the value of the HUBO polynomial at every binary value point immediately before and after the known optimal point. The figure reveals the structure of the space around the optimal point.

Next, we demonstrate that the problem is technically solvable on the current quantum and classical solvers. After translating the HUBO problem into a QUBO problem, we are technically able to utilize hybrid quantum annealers, CPLEX, and Gurobi.

The minimization problem is still extremely hard since we should find the minimum for the holistic optimization problem. On the other hand, since we know the optimum, the problem provides a certain benchmark to compare the systems. D-wave's LeapHybrid is a hybrid solver utilizing classical and quantum annealing computers to divide and solve problems in feasible sized pieces. LeapHybrid solver returned energy values between 449.0 and 236.0 on different optimization rounds. It consistently used either 0.171 seconds or 0.213 seconds of quantum annealing processing time, having a total hybrid running time of around 20 seconds. LeapHybrid solver does not allow the user to modify the solver's hyperparameters (e.g., annealing time) in the annealing process, so we were not able to get better results with it.

We also used another hybrid solver implemented by D-wave, the Kerberos sampler. With Kerberos, we obtained energies between 133.0 and 362.0 on different optimization rounds. Kerberos is a hybrid sampler that utilizes DWaveSampler, simulated annealing, and tabu search. It allows more modifications to the hyperparameters, but we could not find a parameter combination that would improve the results remarkably. With more advanced hybrid methods, it might be possible to obtain better results.

Gurobi and CPLEX performed better in this task. Gurobi reached an energy value of 2 (optimal being 0) and spent around 50 minutes in the optimization process. We stopped the process after Gurobi did not seem to find visible improvements to the best-known solution. CPLEX reached an energy value of 8. All in all, the optimization problem appeared very difficult for modern solvers, although they are technically capable of accepting large QUBO problems.




\section{Discussion}

As far as we know, prior research has not explored the application of quantum computing to tackling tensor-related problems, particularly in computing tensor decompositions. In this study, we develop two innovative algorithms designed specifically for computing tensor decompositions. Furthermore, we show the discovery process of the Strassen algorithm and delve into the characteristics of these algorithms.


One of the big goals of the quantum computing community is to find practical quantum advantage or utility \cite{kim_2023, dwave_supremacy, Arute_Arya_Babbush_Bacon_Bardin_Barends_Biswas_Boixo_Brandao_Buell}. One of the hardest challenges in this task is to identify suitable problem classes that could demonstrate quantum advantage: the problems should not be too easy so that we can solve them efficiently with classical computers, but they should not be too hard for the current, noisy quantum computers with a limited number of qubits and restricted connectivity. The HUBO problems in our decompositional and holistic algorithms contain just thousands of variables (Fig.~\ref{fig:holistic_scalability_1}). Having thousands of variables is not a lot compared to the capabilities of the current quantum annealers since D-Wave claims to support hundreds of thousands of variables. The challenge why classical solvers were not able to solve these problems is because of the high number of interactions between the variables. While the current quantum computers still struggle in encoding large numbers of variables, they are excellent at encoding a vast number of complex interactions. Tensor-based problems may have advantageous characteristics that make them well-suited for quantum computers.

Comparison of quantum algorithms to classical algorithms has proven to be challenging. However, our work introduces a novel perspective through the lens of meta-algorithmic analysis, offering a fresh angle for comparison. When comparing quantum and classical variants of this algorithmic type, the performance criterion is relatively straightforward: the quantum algorithm outperforms its classical counterparts if it can identify a faster matrix multiplication algorithm. Computational time requirements become less critical due to the static and highly complex nature of the problem. In other words, do we need to keep developing faster algorithms to discover the current matrix multiplication algorithms again? The answer is likely no, as accelerating this process would not advance matrix multiplication algorithms themselves.


Our research question of computing tensor decompositions and discovering practical matrix multiplication algorithms is inspired by AlphaTensor \cite{Fawzi_Balog_Huang_Hubert_Romera_Paredes_Barekatain_Novikov_RRuiz_Schrittwieser_Swirszcz_et_2022}, where the authors formulated the problem as a game in which the deep reinforcement learning algorithm aims to select sequences of vectors $\mathbf{x}$, $\mathbf{y}$ and $\mathbf{z}$ so that it would obtain better matrix multiplication algorithms. In this work, we aim to replace deep reinforcement learning with a quantum computer, which provides us with these vector sequences as a solution to binary optimization problems.



We suggest that the decompositional method could be integrated as an alternative short-term reward system for a reinforcement algorithm such as AlphaTensor. Instead of playing a full game by selecting the whole decomposition, the agent in the reinforcement learning algorithm would select only elements $T_{\text{high}}$ and  $\mathbf{x}_{\text{high}} \otimes \mathbf{y}_{\text{high}} \otimes \mathbf{z}_{\text{high}}$ and obtain the tensor decomposition from a quantum computer using the decompositional algorithm.



\section{Conclusion and future work}

In this work, we focused on formulating the problem of finding tensor decompositions for discovering practical matrix multiplication algorithms into a suitable form to be solved on quantum computing hardware. We proposed two algorithms: decompositional and holistic. The decompositional algorithm is able to discover the Strassen algorithm. The holistic algorithm is theoretically capable of finding fixed-length tensor decompositions and, thus, faster matrix multiplication algorithms when the fixed length is shorter than the best-known decomposition. We provided a comprehensive demonstration of these algorithms' functionality, particularly in the context of multiplying $2 \times 2$ matrices. Moreover, we highlighted the specific characteristics of the tensor decomposition problem that make it a promising candidate to be solved on near-future quantum computers.

There are many interesting directions to develop this work further. One direction involves assuming that solutions inherently possess specific symmetries comparable to those outlined in \eqref{eq:symmetries}. Such symmetries will reduce the number of variables within the formulations. For instance, if we assume the symmetry described in \eqref{eq:symmetries}, we would only require half the number of variables. Additionally, investigating the implications of expanding the integer set is a compelling direction of research. Whereas AlphaTensor \cite{Fawzi_Balog_Huang_Hubert_Romera_Paredes_Barekatain_Novikov_RRuiz_Schrittwieser_Swirszcz_et_2022} assumes tensor terms to be selected from integers between -2 and 2, our model facilitates the straightforward encoding of larger integer sets. 

\section*{Acknowledgment}
The author thanks Jukka K. Nurminen, Arianne Meijer - van de Griend, and Ilmo Salmenperä for helpful conversations and Xanadu, BEIT, and Agnostiq for selecting this work as one of the winners of QHack 2023 and for providing in-kind support.

\bibliographystyle{IEEEtran}
\bibliography{IEEEabrv, ref}

\end{document}